\documentclass[11pt,a4paper]{article}
\usepackage{jheppub}
\usepackage{amsmath}
\usepackage[most]{tcolorbox}
\usepackage{dsfont}
\usepackage{ulem}
\usepackage{natbib}
\usepackage{xcolor}
\usepackage[hang,flushmargin]{footmisc}
\usepackage{tikz-cd}
\usepackage{enumitem}
\usepackage{amsfonts,amssymb, amscd,amsmath,latexsym,amsbsy,bm}
\usepackage{stmaryrd}
\usepackage{todonotes}
\usepackage{float}

\usepackage{romannum}

\makeatletter\renewcommand{\@biblabel}[1]{#1.}\makeatother

\newtcolorbox{empheqboxed}{colback=gray!20, 
 colframe=white,
 width=\textwidth,
 sharpish corners,
 top=0mm, 
 bottom=0pt
}

\title{From dual gauge theories to dual spin models }
\author{Mustafa Mullahasanoglu}
\affiliation{
Department of Physics, Bogazici University,\\ 34342 Bebek, Istanbul, Türkiye\\[-0.4cm]

Feza Gursey Center for Physics and Mathematics, Bogazici University,\\ 34684, Kandilli,
Istanbul, Türkiye
}

\emailAdd{mustafa.mullahasanoglu@std.bogazici.edu.tr}

\abstract{
This brief review surveys recent progress driven by the gauge/Yang-Baxter equation (YBE) correspondence. This connection has proven to be a powerful tool for discovering novel integrable lattice spin models in statistical mechanics by exploiting dualities in supersymmetric gauge theories. In recent years, research has demonstrated the use of dual gauge theories to construct new lattice spin models that are dual to Ising-like models.
}


\begin{document}

\maketitle

\section{Introduction}
The study of exactly solvable models represents a significant field of research, yielding innovative techniques and insights for both physics and mathematics \cite{Baxter:1982zz}. A central challenge in this field is the construction of new, exactly solvable models. The most direct approach to this problem is to find new solutions to the fundamental integrability condition, the Yang-Baxter equation (YBE). In the context of two-dimensional lattice spin models, the YBE simplifies to a form known as the star-triangle relation. A further integrability condition, the star-star relation, can be derived from the star-triangle relation. Notably, the star-star relation is more comprehensive, as there exist models whose Boltzmann weights satisfy the star-star relation but not the star-triangle relation.

The gauge/YBE correspondence \cite{Spiridonov:2010em, Yamazaki:2012cp} has led to an increase in new solutions to the star-triangle and star-star relations, accompanied by a remarkably rich conceptual understanding (see also the reviews for more detailed explanations on the gauge side, integrability side, and the explicit solutions \cite{Yagi:2016oum, Gahramanov:2017ysd, Yamazaki:2018xbx}). This modern approach not only enables the systematic investigation of new integrable lattice spin models but also profoundly enriches our understanding, yielding new developments across diverse branches of physics and mathematics.

Beyond models with basic nearest-neighbor interactions, a further class of challenges involves lattice spin models with next-nearest-neighbor or non-planar interactions, as well as models defined on complex lattices like the Kagome lattice. Such models are often studied via duality transformations from the Ising model, where they are solved exactly by mapping their partition functions using symmetry transformations.

This review revisits recent progress in exploring dual lattice spin models derived from Ising-like models discovered through the gauge/YBE correspondence. We will see how dualities in supersymmetric gauge theories not only yield novel integrable models but also provide a framework for understanding their dual counterparts in statistical mechanics.
\section{Integrable models}
We begin by introducing an anisotropic two-dimensional square lattice model. Spin multiplets $\sigma_i=(x_i,m_i)$ sit on the vertices and interact with their nearest neighbors via the edges. In the Ising-like models we consider, these spin multiplets comprise two types of variables: a continuous variable $x \in \mathbb{R}$ and a discrete variable $m \in \mathbb{Z}$. Thus, the multiplet $\sigma_i$ takes values in $\mathbb{R} \times \mathbb{Z}$ or its subset.

The Boltzmann weight for a horizontal edge interaction between nearest-neighbor spins is denoted by $W_{\alpha}(\sigma_i, \sigma_j)$, where $\alpha$ is a spectral parameter that depends on the difference of rapidity variables in this discussion. For vertical interactions, the Boltzmann weight $\overline{W}{\alpha}(\sigma_i, \sigma_j)$ can be related to the horizontal weight via the reflection property $\overline{W}{\alpha}(\sigma_i, \sigma_j) = W_{\eta-\alpha}(\sigma_i, \sigma_j)$ used in this context. Additionally, the models include a self-interaction term $S(\sigma_i)$, which represents the contribution of an external field.

The partition function of these models can be computed exactly using various mathematical techniques \cite{Agirbas:2025xgd}. In this work, we employ the transfer matrix method, where the computation of the partition function reduces to solving an eigenvalue problem for the transfer matrices. Integrability is established by the commutativity of these transfer matrices for different values of the spectral parameter, which implies that they can be simultaneously diagonalized.

The commutativity of transfer matrices is guaranteed by a local integrability condition imposed on the Boltzmann weights, namely the star-triangle relation or the star-star relation. Therefore, demonstrating that a model's Boltzmann weights satisfy one of these relations is sufficient to prove the exact solvability (used as the same as integrability in this discussion) of its partition function.
\subsection{The star-triangle relation}
The star-triangle relation, a fundamental form of the Yang-Baxter equation, ensures the exact computability of the partition function. Consequently, constructing an integrable spin model reduces to finding new solutions to the following equation
\begin{align}
\sum_{m_0} \int dx_0 S(\sigma_0)\prod_{i=1}^3 W_{\alpha_i}(\sigma_i,\sigma_0)
= \mathcal{R}(\alpha_i)
\prod_{1\leq i<j\leq 3}
W_{\alpha_i+\alpha_j}(\sigma_i,\sigma_j)
,\label{str}
\end{align}
where the summation and integration are performed over the central spin $\sigma_0=(x_0,m_0)$, with the spectral parameters constrained by $\sum_{i=1}^3\alpha_i=\eta$. The factor $\mathcal{R}(\alpha_i)$ is a function independent of the spin variables.

Historically, only a handful of solutions to the star-triangle relation were known, and systematic methods for discovering new ones were lacking. The advent of the gauge/YBE correspondence dramatically changed this landscape, leading to the discovery of numerous new solutions. Furthermore, it enabled the construction of a master solution that encompasses and generalizes previously known special cases.

From the perspective of the gauge/YBE correspondence, the star-triangle relation emerges from the equality of partition functions for a pair of dual supersymmetric gauge theories defined on a manifold $\mathcal{M}$
\begin{align}
\mathcal{Z}^{\textit{Theory A}}_{\mathcal{M}}=\mathcal{Z}^{\textit{Theory B}}_{\mathcal{M}}.
\end{align}
Specifically, this duality relates a four-dimensional $\mathcal{N}=1$ (or a three-dimensional $\mathcal{N}=2$) theory (Theory A) with an $SU(2)$ gauge group and $N_f=6$ flavors to theory (Theory B) with fifteen chiral multiplets. In Theory A, the chiral multiplets transform in the fundamental representation of both the gauge and flavor groups, while the vector multiplet transforms in the adjoint representation of the gauge group. In Theory B, the chiral multiplets belong to the totally antisymmetric tensor representation of the flavor group.

By constructing specific dualities of this type, solutions to the star-triangle relation have been obtained in terms of various special functions, including the lens elliptic gamma function \cite{Kels:2015bda}, the elliptic gamma function \cite{Spiridonov:2010em}, the lens hyperbolic gamma function \cite{Gahramanov:2016ilb, Bozkurt:2020gyy, Gahramanov:2022jxz}, the hyperbolic gamma function \cite{Spiridonov:2010em}, trigonometric gamma functions \cite{Gahramanov:2015cva, Catak:2022glx}, and rational gamma functions \cite{Kels:2013ola, Kels:2015bda, Eren:2019ibl}.
\subsection{The star-star relation}
Integrable lattice spin models can also be constructed by solving the star-star relation. The most general solution known to date, which accounts for multi-spin interactions at a single site, was presented in \cite{Yamazaki:2013nra}. 

The star-star relation \cite{Catak:2021coz, Mullahasanoglu:2021xyf} is given by
\begin{align}
 \sum_{m_0}   \int dx_0 S(\sigma_0)\prod_{i=1}^4 W_{\alpha_i}(\sigma_i&,\sigma_0)
  =\frac{W_{2\eta-\alpha_3-\alpha_4}(\sigma_1,\sigma_2)W_{2\eta-\alpha_2-\alpha_3}(\sigma_1,\sigma_4)}{W_{2\eta-\alpha_2-\alpha_3}(\sigma_2,\sigma_3)W_{2\eta-\alpha_3-\alpha_4}(\sigma_3,\sigma_4)}
 \nonumber\\ 
&\times \sum_{m_0}  \int dx_0 S(\sigma_0)\prod_{i=1}^4 W_{\eta-\alpha_i}(\sigma_i,\sigma_0),
    \label{ssr}
\end{align}
where we have a condition on the spectral parameters $\sum_{i=1}^4\alpha_i=\eta$.

From the supersymmetric gauge theory perspective, the star-star relation corresponds to a duality where Theory A has chiral multiplets transforming under the fundamental representation of an $SU(2)$ gauge symmetry and an $SU(8)$ flavor symmetry. Theory B also possesses an $SU(2)$ gauge symmetry, but with a flavor symmetry of $SU(4) \times SU(4) \times U(1)_B \times U(1)_R$. In this theory, the chiral multiplets include a singlet and transform in the fundamental representation of one of the $SU(4)$ factors, while the remaining field content is the same as that of Theory A.
\section{Dual spin models}
The previous section briefly outlined how integrable models are extracted from supersymmetric gauge theories. We now turn to the application of duality transformations to these recently discovered models.

We classify these transformations into two main types. The first type involves modifying the lattice structure itself via decoration transformations and flipping relations, which add or remove spins (added) from a square lattice. As shown in \cite{Fisher1959}, this leads to a duality where the partition functions of the original and decorated models are equal up to a known factor $\mathcal{F}$
\begin{align}
\mathcal{Z}^{\textit{Ising-like~model}}_{\text{Square~lattice}}=\mathcal{Z}^{\textit{Nearest-neighbor~interaction}}_{\text{Decorated~lattice}}\times\mathcal{F}\:.
\end{align}
The second type of transformation maps a model with nearest-neighbor interactions to one with non-planar or next-nearest-neighbor interactions on the same lattice, using relations like the star-square or generalized star-triangle relation. For instance, in the dual Ising model constructed via the star-square relation \cite{Wu1971}, the partition functions are related by
\begin{align}
\mathcal{Z}^{\textit{Ising-like~model}}_{\text{Square~lattice}}=\left(\mathcal{Z}^{\textit{Non-planar~interaction}}_{\text{Square~lattice}}\right)^{\frac{1}{2}}\times\mathcal{G}\:.
\end{align}
\subsection{The decoration transformation }
Integrable lattice spin models can be locally modified or entirely transformed into a different lattice structure provided their Boltzmann weights satisfy the decoration transformation \cite{Catak:2024ygo, Catak:2025fye, Akbulut:2025kow}
\begin{align}
   \sum_{m_0} \int dx_0\: S(\sigma_0) W_{\alpha}(\sigma_1,\sigma_0)W_{\beta}(\sigma_2,\sigma_0)
   =\mathcal{R}(\alpha,\beta) W_{\alpha + \beta}(\sigma_1,\sigma_2)
   \label{dcr},
\end{align}
where the spectral parameters $\alpha$ and $\beta$ are unconstrained.

The decoration transformation can be viewed as a reduced form of the star-triangle relation. Graphically, this corresponds to freezing one spin on the star (removing one leg) and eliminating two edges from the triangle. From the gauge theory perspective, this reduction \cite{Mullahasanoglu:2024stv} is equivalent to integrating out two flavors from the dual theories associated with the star-triangle relation. In this reduced duality, Theory A possesses an $SU(4)$ flavor symmetry, while Theory B contains seven chiral multiplets; all other components of the theories remain unchanged.
\subsection{The flipping relation}
A one-dimensional transformation known as the flipping relation  \cite{Catak:2024ygo, Catak:2025hoz} can be derived as a reduction of the two-dimensional star-star relation  
\begin{align}
\sum_{m_0} \int dx_0 & S(\sigma_0) W_{\alpha}(\sigma_1,\sigma_0)W_{\beta}(\sigma_2,\sigma_0)
\nonumber \\ &= \sum_{m_0} \int dx_0 S(\sigma_0) W_{\beta}(\sigma_1,\sigma_0)W_{\alpha}(\sigma_2,\sigma_0)
\label{flp},
\end{align}
where the spectral parameters $\alpha$ and $\beta$ are unconstrained. The flipping relation differs from other transformations in that it does not alter the lattice structure. Instead, it ensures the invariance of the partition function under an exchange of spectral parameters on adjacent edges, thereby guaranteeing the consistency of applying the decoration transformation in different orders.
\subsection{The star-square relation}
Interestingly, the same duality of gauge theories reduced from the star-triangle can be investigated as the star-square relation \cite{Mullahasanoglu:2023nes}
\begin{align}
  \sum_{m_0} \int dx_0 S(\sigma_0)\prod_{i=1}^4 W(\sigma_i,\sigma_0)
   =
\widetilde{V}(\sigma_1,\sigma_2,\sigma_3,\sigma_4)
   \prod_{1\leq i<j\leq 4} 
   V(\sigma_i,\sigma_j)     
  \:.\label{stsqr}
\end{align}
The star-square relation represents a duality between two distinct types of lattice spin models in statistical mechanics. This transformation maps a configuration with four nearest-neighbor interactions meeting at a central spin (the star) to a configuration without the central spin, featuring seven interactions: four nearest-neighbor interactions $V(\sigma_i,\sigma_j)$ with modified strengths, two next-nearest-neighbor interactions $V$, and one four-spin interaction $\widetilde{V}$ (the square).
\subsection{The generalized star-triangle relation}
The generalized star-triangle relation \cite{Mullahasanoglu:2023nes} transforms a configuration of three nearest-neighbor interactions meeting at a central spin into a new configuration comprising four interactions: three nearest-neighbor interactions with modified strengths and one three-spin interaction
\begin{align}
\begin{aligned}\nonumber
      \sum_{m_0} \int dx_0 S(\sigma_0)\prod_{i=1}^3 W(\sigma_i,\sigma_0) 
   =\Tilde{V}(\sigma_1,\sigma_2,\sigma_3) \prod_{1\leq i<j\leq 3} 
   V(\sigma_i,\sigma_j) \prod_{i=1}^3\overline{S}(\sigma_i)
   \label{gstr},
   \end{aligned}
\end{align}
where $\overline{S}$ is self-interactions for external spins.
Explicit solutions to this relation are obtained from integral identities that arise from the equality of partition functions of dual three-dimensional supersymmetric gauge theories. In this context, a fugacity parameter in the dual theories is fixed by the condition inherited from the previous duality, which is interpreted as the star-square relation.
\section{Conclusions} 
This brief review has outlined a powerful program for constructing integrable and their dual models, where integrable lattice spin models with nearest-neighbor interactions are systematically derived from the duality properties of supersymmetric gauge theories—a connection known as the gauge/YBE correspondence. We have highlighted how this correspondence not only generates new solutions to fundamental integrability conditions but also provides a unified framework for understanding duality transformations between different spin models. The exploration of these connections, facilitated by the rich mathematical structure of special functions and symmetries, promises to remain a fertile ground for future research.
\section*{Acknowledgements} 
I am deeply grateful to Ilmar Gahramanov, Erdal Catak and all my collaborators from the Istanbul Integrability and Stringy Topics Initiative (\href{https://istringy.org/}{istringy.org}) for their contributions to the research that forms the basis of this review. This work is based on talks presented at the department seminar of the University of Tokyo, at the Central European Institute for Cosmology and Fundamental Physics (CEICO) in Prague, and at the Quantum Theory and Symmetries (QTS-13) Symposium in Yerevan, 2025. 

I extend my sincere thanks to Masahito Yamazaki and Joris Raeymaekers for their warm hospitality during my visits to the Kavli Institute for the Physics and Mathematics of the Universe (IPMU) and the CEICO, respectively. This work was supported by the Scientific and Technological Research Council of Turkey (TÜBİTAK) under the grants with numbers 122F451 and 123N952.
\bibliographystyle{utphys}
\bibliography{refYBE}

\end{document}